\documentclass[aps,prl,showpacs,amsmath,amssymb,superscriptaddress,reprint,10pt]{revtex4-1}
\usepackage{bm}
\usepackage{graphicx}
\usepackage{xcolor}
\usepackage{url}
\usepackage[colorlinks=true,breaklinks=true,allcolors=blue]{hyperref}

\newcommand\rmd{{\mathrm{d}}}
\newcommand\rme{{\mathrm{e}}}
\newcommand\rmi{{\mathrm{i}}}
\DeclareMathOperator{\Tr}{Tr}

\begin{document}

\title{Universal Threshold for the Dynamical Behavior of Lattice Systems with Long-Range Interactions}  

\author{Romain Bachelard} 
\email{bachelard.romain@gmail.com} 
\affiliation{Instituto de F\'{\i}sica de S\~ao Carlos, Universidade de S\~ao Paulo, 13560-970 S\~ao Carlos, SP, Brazil}

\author{Michael Kastner} 
\email{kastner@sun.ac.za} 
\affiliation{Institute of Theoretical Physics,  University of Stellenbosch, Stellenbosch 7600, South Africa}
\affiliation{National Institute for Theoretical Physics (NITheP), Stellenbosch 7600, South Africa} 

\date{\today}

\begin{abstract}
Dynamical properties of lattice systems with long-range pair interactions, decaying like $1/r^\alpha$ with the distance $r$, are investigated, in particular the time scales governing the relaxation to equilibrium. Upon varying the interaction range $\alpha$, we find evidence for the existence of a threshold at $\alpha=d/2$, dependent on the spatial dimension $d$, at which the relaxation behavior changes qualitatively and the corresponding scaling exponents switch to a different regime. Based on analytical as well as numerical observations in systems of vastly differing nature, ranging from quantum to classical, from ferromagnetic to antiferromagnetic, and including a variety of lattice structures, we conjecture this threshold and some of its characteristic properties to be universal.
\end{abstract}


\maketitle 

Most lattice models studied in condensed matter phys\-ics have interactions of finite range. Despite the long-range character of the fundamental electromagnetic interactions, the presence of positive and negative charges gives rise to screening effects that cause interactions to be effectively short range and justify an approximation by finite-range (and often even nearest-neighbor) interactions. But a finite-range approximation is not always justified. One obvious example are astrophysical systems dominated by gravitational interactions where screening does not occur. Historically, it was in this context that several of the anomalies of long-range interacting systems were first discussed, for example negative heat capacities of bounded self-gravitating gas spheres \cite{LynWood68,*Thirring70}. In subsequent decades, a general interest in fundamental issues of statistical physics of long-range interacting systems arose \cite{CamDauxRuf09}. 

Technically, interactions of finite range turn out to be convenient: Exact solutions of many-body models, rare as they are, are in most cases restricted to finite-range interactions, and also general theorems in thermostatistical physics (like the convexity properties of thermodynamic functions) are often proved under the assumption of short-range interactions \cite{Ruelle}. ``Short-range'' here refers not only to interactions of finite range, but also to those decaying like $1/r^\alpha$ with the distance $r$ and an exponent $\alpha$ larger than the spatial dimension $d$ of the system. And indeed, various long-range systems, i.e., those with exponents $\alpha<d$, have been found to violate ensemble equivalence, have negative microcanonical specific heat, and other peculiarities \cite{CamDauxRuf09}.

Out of equilibrium, an intriguing and physically relevant phenomenon that has been observed in long-range systems is the existence of quasistationary states. These are nonequilibrium states whose lifetimes $\tau$ diverge with increasing system size $N$. As a result, for a sufficiently large system, relaxation to equilibrium takes place on a time scale that is larger than any realistic observation time. Such behavior has been reported for various long-range systems, ranging from classical toy models with mean-field interactions to gravitating systems \cite{AnRu95,*JoyceWorrakitpoonpon10}. The exponent $\alpha$ in these studies is usually kept fixed---either at $\alpha=0$ (corresponding to mean-field interactions) where analytical calculations are easier, or at some integer number in order to account for classical gravity. 

\begin{figure}
{\center
\includegraphics[width=0.48\linewidth]{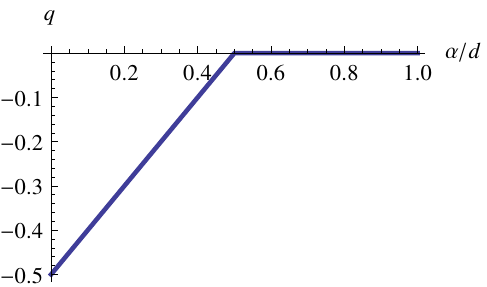}
\includegraphics[width=0.48\linewidth]{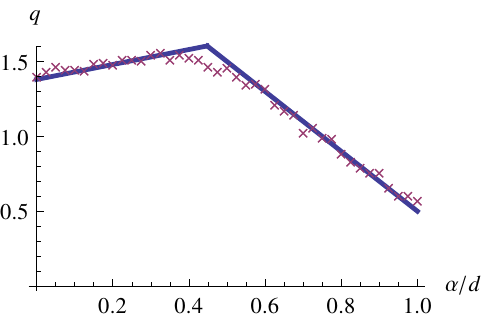}
}
\caption{\label{f:scaling}%
(Color online) The exponent $q$ of the scaling law $\tau\propto N^q$ that governs the system-size dependence of the relaxation time $\tau$. Left: For the long-range quantum Ising model \eqref{e:H} the exponent $q=\min\{0,\alpha/d-1/2\}$ follows from the bounds in~\eqref{e:bound}. Right: For the $\alpha XY$ chain \eqref{e:HMF}. The crosses mark data points, obtained by a scaling analysis as in Fig.~\ref{f:kurtosis}. The line is plotted as a guide for the eye, indicating two distinct linear regimes with a crossover at $\alpha/d\approx0.44$.
}%
\end{figure}

\begin{figure}
{\center
\includegraphics[width=0.48\linewidth]{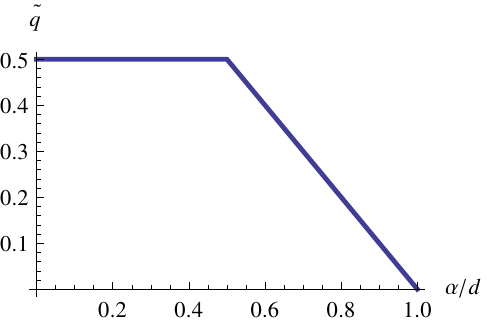}
\includegraphics[width=0.48\linewidth]{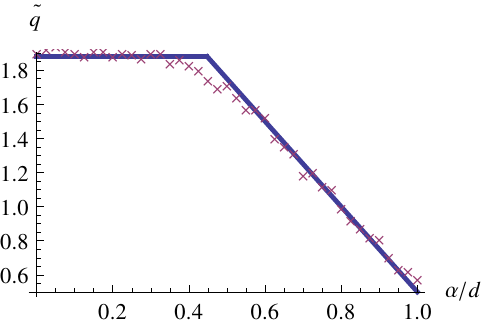}
}
\caption{\label{f:scalingtilde}%
(Color online) As in Fig.\ \ref{f:scaling}, but for scaling exponents $\tilde{q}$ as modified by the presence of an $N$-dependent prefactor $\mathcal{N}$ in the Hamiltonian. The left plot is for the long-range quantum Ising model, the right for the classical $\alpha XY$ chain.
}%
\end{figure}

In this Letter we investigate the relaxation to equilibrium of lattice systems with long-range pair interactions, and in particular the $\alpha$ dependence of the relaxation times. We report analytical as well as numerical results on several classes, containing systems of vastly differing nature. For all lattice structures studied, and regardless of whether the systems are quantum or classical, we observe, at a threshold value of $\alpha=d/2$, a drastic change of the relaxational dynamics. In particular, the exponent $q(\alpha)$, governing the scaling $\tau\propto N^{q(\alpha)}$ of the relaxation time $\tau$, switches from one regime to another at $\alpha=d/2$. Certain qualitative aspects of the scaling laws also appear to be universal, as summarized in Figs.\ \ref{f:scaling} and \ref{f:scalingtilde}. Note that the threshold value $\alpha=d/2$ that is relevant for nonequilibrium phenomena differs from the one at $\alpha=d$ commonly used for distinguishing between long- and short-range behavior in equilibrium statistical mechanics. Other aspects of the dynamics are evidently nonuniversal, even to the point that relaxation in some systems may slow down with increasing system size, but accelerate in others.

This Letter aims at furthering the understanding of fundamental aspects of nonequilibrium statistical mechanics of long-range interacting systems. In particular, the mechanism of relaxation in long-range interacting systems, but also many aspects of its phenomenology, are still only poorly understood. Identifying universal properties and threshold values may give valuable clues and deepen the general understanding. On the more applied side, recent developments have made long-range interacting systems accessible in the laboratory, and an experimental verification of some of our results should be feasible~\cite{ODell_etal00,*Dominguez_etal10,*Golestanian12,*Chalony_etal12}. Particularly promising for such a check is an ion-trap-based quantum simulator as reported by Britton {\em et al.}\ \cite{Britton_etal12}. In this setup, a quantum long-range Ising model is emulated, and the exponent $\alpha$ governing the interaction range can be tuned between 0 and 3.

{\em Long-range quantum Ising model.---}The first class of models we are studying consists of spin-$1/2$ degrees of freedom on the $N$ sites of a lattice $\Lambda$, governed by the Hamiltonian operator
\begin{equation}\label{e:H}
H_\Lambda=-\frac{J}{2}\sum_{\substack{i,j\in\Lambda\\i\neq j}}\frac{\sigma_i^z\sigma_j^z}{|i-j|^\alpha} - h\sum_{i\in\Lambda}\sigma_i^z.
\end{equation}
Here, $\sigma^z$ is the $z$ component of the Pauli spin operator, $h$ is an external magnetic field, $|i-j|$ denotes the Euclidean distance of sites $i$ and $j$, and the exponent $\alpha>0$ determines the interaction range. Depending on the sign of the coupling $J$, the spin--spin interactions are ferromagnetic ($J>0$) or antiferromagnetic ($J<0$). On certain lattices (e.g., a triangular lattice), antiferromagnetic interactions result in geometrical frustration. An important difference to a similar model considered in \cite{Kastner11,*Kastner12} is the absence of an $N$-dependent normalization factor in front of the first sum in \eqref{e:H}, introduced in \cite{Kastner11,*Kastner12} to ensure extensivity of the energy. We will comment on the role of the factor later in this Letter.

The expectation value $\langle A\rangle$ of an observable $A$ is given by $\Tr[A\rho(t)]$, where the time evolution of the density operator $\rho$ is governed by the von Neumann equation. For all initial density operators $\rho_0$ that are diagonal in the $\sigma^x$ tensor product eigenbasis, the time evolution of single-spin observables can be computed analytically in arbitrary spatial dimension, yielding
\begin{equation}\label{e:exact}
\langle\sigma_i^x\rangle(t)=\langle\sigma_i^x\rangle(0)\cos(2ht)\prod_{\substack{j\in\Lambda\\j\neq i}} \cos \left(\frac{2Jt}{|i-j|^\alpha}\right)
\end{equation}
(see Supplemental Material A). For all finite lattices $\Lambda$, \eqref{e:exact} is a quasiperiodic function. Proper relaxation to equilibrium can be observed in the thermodynamic limit of infinite lattice size, where using techniques analogous to those in \cite{Emch66,*Radin70,Kastner11,*Kastner12,vandenWorm_etal}, we obtain
\begin{multline}\label{e:bound}
\left|\langle\sigma_i^x\rangle(t)\right|\leq\left|\langle\sigma_i^x\rangle(0)\cos(2ht)\right|\\
\times
\begin{cases}
\exp\left[-\frac{2^{5+2\alpha-d}\pi^{d/2-2}}{(d-2\alpha)\Gamma(d/2)}J^2 t^2 N^{1-2\alpha/d}\right] & \text{for $0\leq\alpha<d/2$},\\[2mm]
\exp\left[-\frac{2^{1+d-2\alpha}\pi^{d/2}}{(2\alpha-d)\Gamma(d/2)}\left\lvert\frac{4Jt}{\pi}\right\rvert^{d/\alpha}\right] & \text{for $\alpha>d/2$},
\end{cases}
\end{multline}
valid for large $N$ and $t$ (see Supplemental Material B). This result differs from the one reported in \cite{Kastner11,*Kastner12} in a nontrivial way. This is a consequence of the absence of the above-mentioned $N$-dependent normalization in the Hamiltonian \eqref{e:H}, which essentially corresponds to taking the limits of large $N$ and $t$ in a different way. The functional form of the upper bounds in \eqref{e:bound} indeed correctly reflects the behavior of $\left|\langle\sigma_i^x\rangle(t)\right|$, i.e., the powers of $N$ and $t$ in the exponent agree excellently with a numerical evaluation of \eqref{e:exact} for large $N$, and only the numerical constants are, as expected, overestimated (Supplemental Material B).

From Eq.\ \eqref{e:bound} it follows that a change of regime takes place at $\alpha=d/2$. For $\alpha<d/2$, relaxation to equilibrium is Gaussian in time, and the corresponding relaxation time $\tau$ scales like $\tau\propto N^{\alpha/d-1/2}$, shrinking to zero in the limit of large $N$. For $\alpha>d/2$, relaxation is governed by a compressed or stretched exponential in $t$, with a relaxation time that is constant asymptotically for large $N$. These scaling laws are summarized in Fig.\ \ref{f:scaling} (left). The threshold at $\alpha=d/2$ suggests the following interpretation: Only for $\alpha<d/2$ are the pair interactions sufficiently long-range such that the relaxation dynamics of a single spin is directly influenced, and thereby sped up, by the presence of $\mathcal{O}(N)$ spins. Analytical calculations of spin--spin correlation functions indicate that further significant qualitative and quantitative changes occur in other dynamical quantities \cite{vandenWorm_etal}. Interestingly, this dynamical long-range threshold differs from the equilibrium threshold at $\alpha=d$ below which peculiar long-range behavior, like nonequivalence of ensembles or negative specific heat, may occur in equilibrium statistical physics.

{\em $\alpha XY$ chain.---}This model, introduced in \cite{AnteneodoTsallis98}, consists of classical $XY$ spins attached to the sites $i$ of a one-dimensional chain and parametrized by the angular variables $\phi_i$. The time evolution is generated, via Hamilton's equations, by the Hamiltonian function
\begin{equation}\label{e:HMF}
H(p,\phi)=\sum_{i=1}^N\frac{p_i^2}{2} - \frac{J}{2}\sum_{\substack{i,j=1\\i\neq j}}^N \frac{\cos(\phi_i-\phi_j)}{|i-j|^\alpha},
\end{equation}
where $\phi=(\phi_1,\dotsc,\phi_N)$ is the vector of angle variables and $p=(p_1,\dotsc,p_N)$ the vector of conjugate momenta. For $\alpha=0$, and besides a normalization factor $1/N$ in front of the second sum, Eq.\ \eqref{e:HMF} reduces to the much studied Hamiltonian Mean-Field model \cite{AnRu95}, a model known to display many of the peculiarities of long-range interacting systems. In particular, relaxation to equilibrium has been studied extensively, and the occurrence of quasistationary states was observed for large classes of initial conditions \cite{LatoraRapisardaTsallis01,*Yamaguchi_etal04,*CagliotiRousset08}. In equilibrium and for exponents $0\leq\alpha<1$, the model \eqref{e:HMF} shows a transition from a magnetized phase at energy densities $e=E/N<J/4$ to an unmagnetized phase for $e>J/4$ \cite{GiansantiMoroniCampa02}.

Initially we prepare the system in so-called waterbag initial distributions, with initial angles $\phi_i$ drawn from a flat distribution, and initial momenta taking random values in the symmetric interval $[-\Delta,+\Delta]$ with some $\Delta>0$; the average energy per particle is then $e=\Delta^2/6$. The time evolution is investigated by numerically integrating the Hamiltonian equations of motion, using a sixth-order symplectic integrator \cite{McLachlanAtela92}. In earlier studies of the $\alpha=0$ case, the magnetization $m$ had been monitored over time in order to observe relaxation to equilibrium \cite{Yamaguchi03}. This approach is not viable above the critical energy $e_\text{c}=1/4$ where the initial and the equilibrium value of $m$ are both close to zero. For that reason, we monitor the time evolution of the kurtosis of the momentum distribution, $\kappa=\langle p^4\rangle/\langle p^2\rangle^2$, where the angular brackets denote averages over the lattice. The kurtosis of the waterbag initial states we are using is $\kappa_0=9/5$, whereas the Boltzmann equilibrium distribution has $\kappa_\text{eq}=3$.

\begin{figure}
{\center
\includegraphics[width=0.48\linewidth]{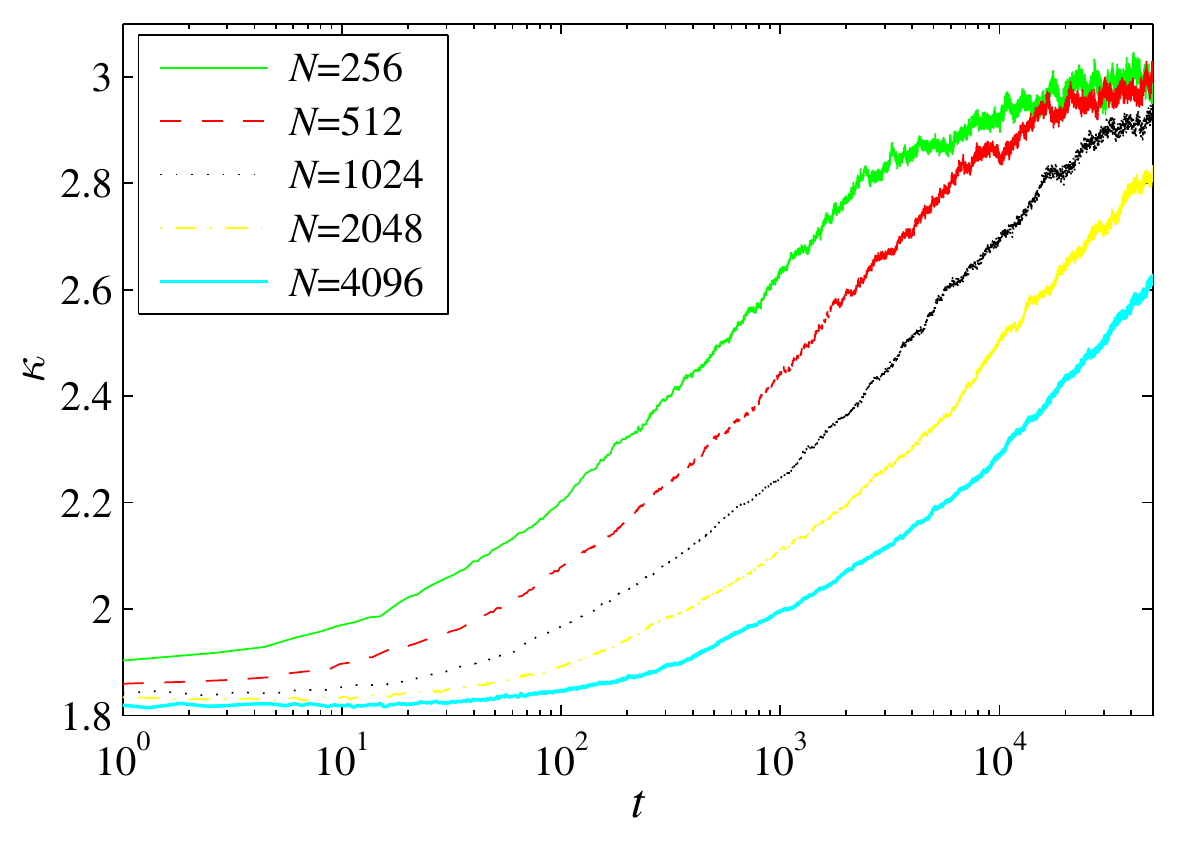}
\includegraphics[width=0.48\linewidth]{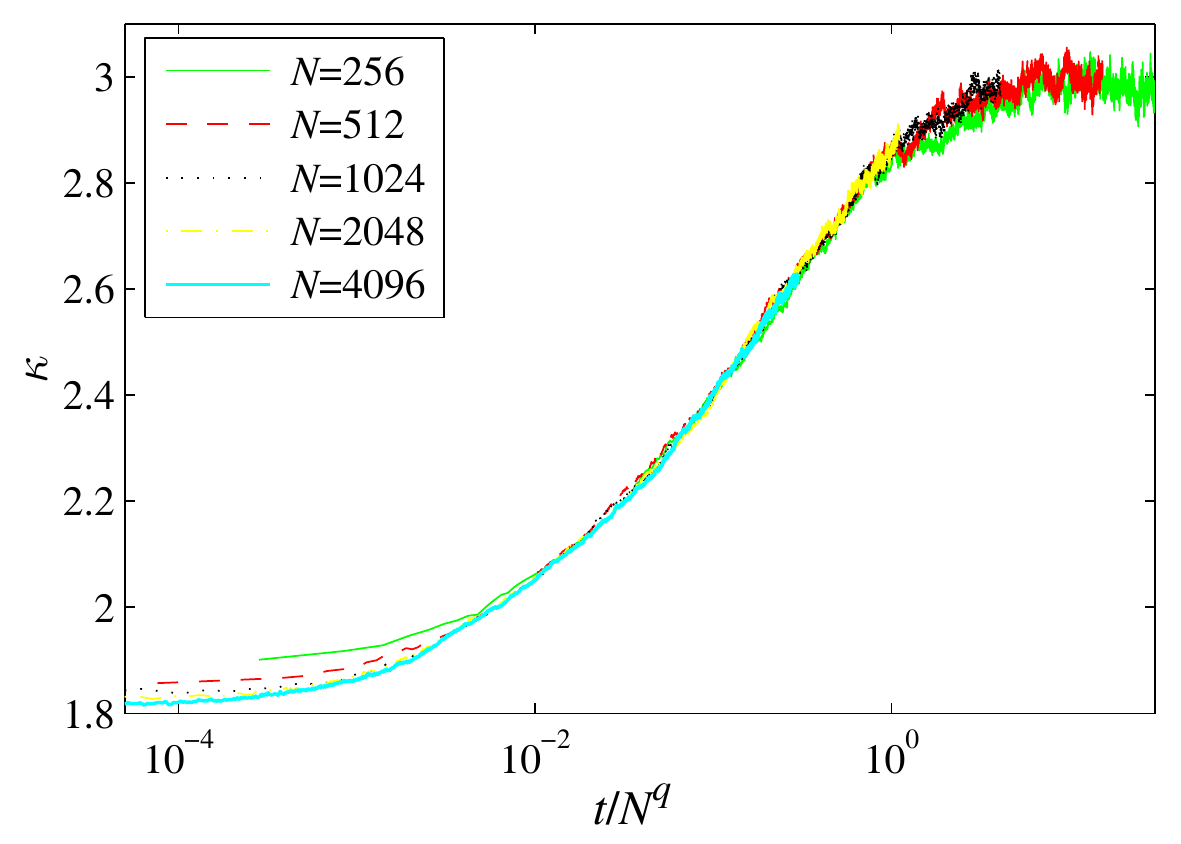}
}
\caption{\label{f:kurtosis}%
(Color online) The kurtosis $\kappa$ of the momentum distribution of the $\alpha XY$ chain, plotted for $\alpha=1/8$, energy density $e\approx0.26$, and system sizes $N=256$, $512$, $1024$, $2048$ and $4096$. Data are averaged over $128$, $64$, $32$, $16$ and $8$ realizations, respectively. Left: As a function of time $t$, the relaxation time increases with system size $N$. Right: As a function of rescaled time $t/N^q$ with scaling exponent $q=1.453$.
}%
\end{figure}

Choosing $\Delta=5/4$, we prepare the system at an energy density $e=25/96\approx0.26$ slightly above the transition energy $e_\text{c}$ in the unmagnetized regime. The time evolution of the kurtosis $\kappa$ is shown in Fig.\ \ref{f:kurtosis} (left) for $\alpha=1/8$. The data reveal a relaxation towards the equilibrium value $\kappa_\text{eq}=3$, on a time scale that depends strongly on the system size $N$. Plotting the same data vs rescaled time $t/N^q$ with $q=1.453$, the curves for different $N$ collapse onto each other, demonstrating the validity of the scaling law; see Fig.\ \ref{f:kurtosis} (right). Performing such a scaling analysis for different values of $\alpha$, we obtain the scaling exponent $q$ as a function of $\alpha$ as shown in Fig.\ \ref{f:scaling} (right). The plot reveals two regimes: For $0\leq\alpha\leq\alpha_\text{th}$ with $\alpha_\text{th}\approx0.45$, $q$ evolves linearly in $\alpha$ as $q\approx1.38+0.5\alpha$. The second regime, again linear in $\alpha$, is described by $q\approx2.5-2\alpha$ for $\alpha_\text{th}\leq\alpha<1$. This confirms, similar to our findings for the quantum Ising model, the presence of two distinct power-law regimes for the relaxation times of the $\alpha XY$ chain. For $\alpha>1$, rescaling of time does not any longer lead to a data collapse, so either no such scaling law exists in this regime, or larger system sizes would be necessary to reach the scaling regime. For other values of the energy density $e$, the data collapse is of lesser quality, but the behavior of $q(\alpha)$ is similar to the one shown in Fig.\ \ref{f:scaling}, though with larger fluctuations (see Supplemental Material C for simulation data and a more detailed discussion).

{\em Normalization of the energy scale.---}In many papers on long-range interacting systems, the pair-interaction term in the Hamiltonian is made extensive by equipping it with an $N$-dependent normalization factor $\mathcal{N}$. For the models studied in this Letter, such normalization factors, as discussed for example in \cite{Kastner11,*Kastner12} and \cite{AnteneodoTsallis98,GiansantiMoroniCampa02}, behave as $\mathcal{N}\sim N^{\alpha/d-1}$ asymptotically for large $N$ when $\alpha<d$ (while no normalization is required for $\alpha>d$). Extensivity of the energy is a prerequisite for a well-defined thermodynamic limit when making the transition from equilibrium statistical mechanics to thermodynamics \cite{Ruelle}. Moreover, interesting physical phenomena, like equilibrium phase transitions, are caused by competition between energetic and entropic effects, and extensivity of both these quantities is usually required in order for such a competition to survive in the thermodynamic limit.

For studying the approach to equilibrium, a normalization factor is not necessary, but can be included \footnote{Care must be taken when performing the limits of large $N$ and $t$, as the prefactor $\mathcal{N}$ may affect the order of these limits.}. An $N$-dependent prefactor in the energy induces a further $N$ dependence of the time scale, shifting the scaling exponent $q$ of the relaxation time scale to $\tilde{q}=q+(1-\alpha/d)$ in quantum dynamics, and to $\tilde{q}=q+(1-\alpha/d)/2$ in classical Hamiltonian dynamics. The dependence of the exponent $\tilde{q}$ on $\alpha$ is shown in Fig.\ \ref{f:scalingtilde} for the long-range quantum Ising model and the classical $\alpha XY$ chain. The similarities between the two models become even more evident in this plot, with a crossover from a constant regime for $\alpha<\alpha_\text{th}$ to a linear regime above the threshold.

In the case of the long-range quantum Ising model, introducing the normalization factor $\mathcal{N}$ has the effect of even flipping the sign of the exponent, from a negative $q$ to a positive $\tilde{q}$. This implies that, while the relaxation to equilibrium takes longer and longer with increasing system size in the presence of $\mathcal{N}$, the opposite happens for the original Hamiltonian: Relaxation speeds up with increasing $N$, leading to ``instantaneous'' equilibration in the thermodynamic limit. This is different from what is observed for the $\alpha XY$ chain, where long-lived quasistationary states occur independently of whether the factor $\mathcal{N}$ is present or not.

{\em Discussion of the results.---}As illustrated in Figs.\ \ref{f:scaling} and \ref{f:scalingtilde}, we find that the exponent $q$ in the scaling law $\tau\propto N^q$ of the relaxation time varies linearly as a function of $\alpha$, with a change from one linear regime to another at $\alpha=\alpha_\text{th}$. For the quantum Ising model, $\alpha_\text{th}=d/2$ is the exact location of this change, whereas for the $\alpha XY$ chain we find an approximate value of $\alpha_\text{th}\approx0.45$. However, results on the largest Lyapunov exponent of the $\alpha XY$ model on $d$-dimensional lattices \cite{FirpoRuffo01,*AnteneodoVallejos01} support the conjecture that the exact value of $\alpha_\text{th}$ is $d/2$ also in the classical case: Lyapunov exponents are characteristic quantities for the time evolution, quantifying in some sense the chaoticity of the dynamics. The authors of \cite{FirpoRuffo01,*AnteneodoVallejos01} find that the largest Lyapunov exponent vanishes like $N^{-\kappa}$, where $\kappa$ changes from a constant regime for $0<\alpha/d<1/2$ to a linearly decreasing regime for $1/2<\alpha/d<1$. The qualitative similarity of $\kappa(\alpha)$ in Fig.\ 3 of \cite{FirpoRuffo01} to the plot of $\tilde{q}(\alpha)$ in Fig.\ \ref{f:scalingtilde} of the present Letter is striking. At least sufficiently far from equilibrium, the sum of positive Lyapunov exponents is known to correspond to an entropy rate, which in turn provides a link to the speed at which equilibrium is approached \cite{LatoraBaranger99}. When the largest term of that sum switches to a different scaling regime, this change is expected to reflect also in the sum and, therefore, in the speed at which the system relaxes to equilibrium. These arguments suggest that the transition from one linear regime of $q(\alpha)$ to another takes place precisely at $\alpha_\text{th}=d/2$, not only for the quantum Ising model, but also for the $\alpha XY$ chain, and our numerical results are in good agreement with this analytical prediction.

Our understanding of the origin of the universality of the threshold at $\alpha_\text{th}=d/2$ is partial at best, but some physical intuition can be gained from studying Lieb-Robinson-type bounds on the propagation of perturbations. Hastings and Koma \cite{HastingsKoma06} report such a bound for a broad class of quantum lattice systems with long-range interactions. One restriction on the interactions (Eq.\ (2.3) of \cite{HastingsKoma06}) is essential for obtaining a nontrivial bound, and it roughly amounts to requiring that
\begin{equation}\label{e:condition}
\sum_{k\in\Lambda}\frac{1}{|i-k|^\alpha}\frac{1}{|j-k|^\alpha} < \infty
\end{equation}
for any given lattice sites $i,j\in\Lambda$. By integral approximation one finds that \eqref{e:condition} is satisfied for $\alpha>d/2$, reproducing the threshold value we found for the relaxation dynamics. This suggests an appealing, though speculative, intuitive explanation: In the regime $\alpha>\alpha_\text{th}$, restrictions on the speed at which perturbations propagate (as given by the Lieb-Robinson bound) are responsible for one type of relaxation behavior, whereas the absence of such restrictions gives rise to another type of relaxation in the regime $\alpha<\alpha_\text{th}$. However, to turn this intuition into a proof of the universality of the threshold at $\alpha_\text{th}$, important pieces are still missing, one of them being a classical version of the Lieb-Robinson bound for long-range interacting systems.

In summary, substantial analytical as well as numerical evidence has been found for the existence of a threshold at $\alpha_\text{th}=d/2$ at which dynamical properties of long-range interacting systems show an abrupt change from one regime to another. This threshold is found for classical Hamiltonian as well as quantum dynamics, on lattices of arbitrary dimension and for various lattice structures, and for ferromagnetic as well as anti-ferromagnetic interactions. Anecdotal evidence that the same threshold value is of significance also in random models is reported in Ref.\ \cite{Wreszinski10}, although here different kinds of randomness may behave differently and the scenario becomes more involved \footnote{This is similar to what happens in equilibrium statistical mechanics, where the threshold at which equilibrium properties switch from long-range to short-range behavior is universal in the absence of randomness, but can get modified when randomness is present.}. Furthermore, a change of regime can be observed not only in the relaxation times we have studied in the present letter, but also in other dynamical quantities, be it the above-mentioned largest Lyapunov exponents \cite{FirpoRuffo01,*AnteneodoVallejos01} or the emergence of widely separated time scales of decaying correlations \cite{vandenWorm_etal}. Beyond the universality of the threshold value $\alpha_\text{th}=d/2$, it is tempting to conjecture on the basis of Fig.\ \ref{f:scalingtilde} that also the two linear regimes in $q(\alpha)$ or $\tilde{q}(\alpha)$ may be model-independent and universal. Admittedly, this is speculation, and understanding the origin of the observed universality still poses challenges for future research. An experimental verification for more general quantum Ising models (i.e., in the presence of a transverse magnetic field) should be possible with the ion trap technology of Ref.\ \cite{Britton_etal12}. 

The authors thank F.\ Bouchet, S.\ Ruffo, F.\ Staniscia, and M.\ van den Worm for fruitful discussions.
M.\,K.\ acknowledges support by the Incentive Funding for Rated Researchers programme of the National Research Foundation of South Africa, and  R.\,B.\ from the Funda\c{c}\~ao de Amparo \`a Pesquisa do Estado de S\~ao Paulo (FAPESP).


\pagebreak

\begin{center}
{\bf Supplemental Material}  
\end{center}
\vspace{-3mm}

\appendix



\section{A.\hspace{2mm} Derivation of equation (2)}

The derivation is similar to, but simpler than, the one reported in Appendix A of \cite{vandenWorm_etal}. Starting point of the calculation is the general expression for the expectation value
\begin{equation}\label{e:spinspin}
\left\langle\sigma_i^\pm\right\rangle\!(t)=\Tr \left(\rme^{\rmi H_\Lambda t} \sigma_i^\pm \rme^{-\rmi H_\Lambda t}\rho_0\right)
\end{equation}
with respect to the initial density operator $\rho_0$, where we assume $\rho_0$ to be diagonal in the $\sigma_x$ tensor product eigenbasis. Spin ladder operators are defined as $\sigma^\pm=(\sigma^x\pm\rmi\sigma^y)/2$. Since all terms in the Hamiltonian $H_\Lambda$ commute, we can factorize the time evolution operator,
\begin{equation}\label{e:U}
\exp\left(-\rmi H_\Lambda t\right)=\prod_{k<l}\exp\left(\rmi J_{k,l}\sigma_k^z\sigma_l^z t\right)\prod_m\exp\left(\rmi h\sigma_m^z t\right),
\end{equation}
and similarly for the Hermitian conjugate. Here we have introduced the notation $J_{i,j}=J/\lvert i-j\rvert ^\alpha$ for the distance-dependent couplings. All factors in \eqref{e:U} that do not contain $\sigma_i^z$ commute with $\sigma_i^\pm$. To compute the time evolution in \eqref{e:spinspin}, we therefore have to deal with the expression
\begin{multline}\label{e:ssoft}
\prod_{k\neq i}\exp\left(-\rmi J_{k,i}\sigma_k^z\sigma_i^z t\right)\exp\left(-\rmi h\sigma_i^z t\right)\sigma_i^\pm\\
\times\exp\left(\rmi h\sigma_i^z t\right)\prod_{l\neq i}\exp\left(\rmi J_{l,i}\sigma_l^z\sigma_i^z t\right).
\end{multline}
Making use of $\left[\sigma^z,\sigma^\pm\right]=\pm2\sigma^\pm$, the time evolution due to the magnetic field $h$ simplifies to
\begin{equation}\label{e:Bevolv}
\exp\left(-\rmi h\sigma_i^z t\right)\sigma_i^\pm\exp\left(\rmi h\sigma_i^z t\right)=\sigma_i^\pm\exp\left(\mp2\rmi ht\right).
\end{equation}
Picking one lattice site $k\neq i$, the time evolution of $\sigma_i^\pm$ due to the interaction with the spin at $k$ can be written as
\begin{multline}\label{e:Jevolv}
\exp\left(-\rmi J_{k,i}\sigma_k^z\sigma_i^z t\right)\sigma_i^\pm
\exp\left(\rmi J_{k,i}\sigma_k^z\sigma_i^z t\right)\\
=\sigma_i^\pm\cos\left(2tJ_{i,k}\right)+\rmi\sigma_i^\pm\sigma_k^z\sin\left(2tJ_{i,k}\right).
\end{multline}
Since the initial state $\rho_0$ is assumed to be diagonal in the $\sigma_x$ tensor product eigenbasis, only diagonal elements of the operator $\rme^{\rmi H_\Lambda t} \sigma_i^\pm\rme^{-\rmi H_\Lambda t}$ in the same basis contribute to the trace in \eqref{e:spinspin}. For this reason we can drop the second term on the right-hand side of \eqref{e:Jevolv}, as it is proportional to $\sigma_k^z$. Inserting \eqref{e:Bevolv} and \eqref{e:Jevolv} into \eqref{e:spinspin}, we obtain
\begin{equation}\label{e:pmpmfinal}
\left\langle\sigma_i^\pm\right\rangle\!(t)=\left\langle\sigma_i^\pm\right\rangle\!(0)\exp\left(\mp2\rmi ht\right)
\prod_{k\neq i}\cos\left(2tJ_{i,k}\right),
\end{equation}
where $\left\langle\sigma_i^\pm\right\rangle\!(0)=\Tr\left(\sigma_i^\pm\rho_0\right)$. From \eqref{e:pmpmfinal} we obtain
\begin{align}
\left\langle\sigma_i^x\right\rangle\!(t)&=\left\langle\sigma_i^-\right\rangle\!(t)+\left\langle\sigma_i^+\right\rangle\!(t)\nonumber\\
&=\left\langle\sigma_i^x\right\rangle\!(0)\cos(2ht)\prod_{k\neq i}\cos\left(2tJ_{i,k}\right).\label{e:xfinal}
\end{align}

\section{B.\hspace{2mm} Derivation of equation (3)}

This derivation shares some similarities with the ones reported in Sec.\ 4.1 of \cite{Kastner12} and Appendix B of \cite{vandenWorm_etal}. The aim is to construct, in the thermodynamic limit of infinite lattice size, a nontrivial upper bound on the product
\begin{equation}\label{e:P_N}
P_\Lambda=\prod_{k\in\Lambda\setminus i} \left\lvert \cos\left(\frac{2Jt}{\lvert i-k\rvert ^\alpha}\right)\right\rvert 
\end{equation}
occurring in the expression for the modulus of the expectation value $\left\langle\sigma_i^x\right\rangle\!(t)$ in \eqref{e:xfinal}. Without loss of generality we set $i=0$ in the following.

\begin{figure}\centering
\includegraphics[width=0.45\linewidth]{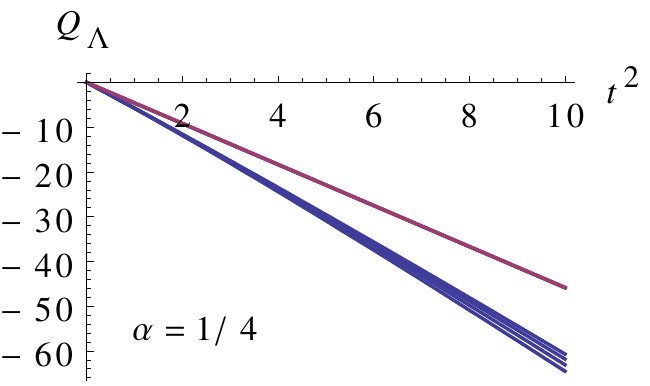}
\includegraphics[width=0.53\linewidth]{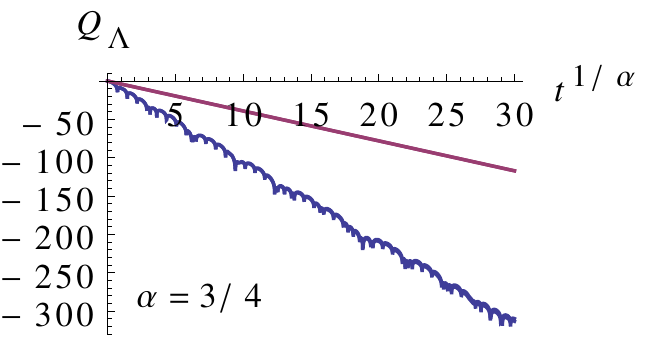}
\caption{\label{f:bounds}%
Comparison of the exact result \eqref{e:P_N} to the upper bound \eqref{e:upperbound}. Left: For $\alpha=1/4$ the rescaled logarithm $Q_\Lambda=N^{2\alpha/d-1}\ln P_\Lambda$ of $P_\Lambda$ is plotted as a function of $t^2$ (by straightforward evaluation of \eqref{e:P_N} with {\sc Mathematica}). With this rescaling, the logarithm of the upper bound in \eqref{e:upperbound} is given by the red (upper) straight line in the left plot. $Q_\Lambda$, evaluated for chains of lengths $N=25001$, 50001, 100001, and 200001 with open boundary conditions, is shown in blue (from top to bottom). All these curves show a linear decay, which confirms that the Gaussian decay $\propto\exp(-ct^2)$ predicted by \eqref{e:upperbound} correctly captures the long-time asymptotics of $P_\Lambda$, but overestimates, as expected, the constant $c$. The fact that the exact results for $Q_\Lambda$ do not collapse onto a single curve indicates, however, that the power of $N$ in the bound \eqref{e:upperbound} is not the sharpest possible one. Right: For $\alpha=3/4$, the logarithm $Q_\Lambda=\ln P_\Lambda$ (without any $N$-scaling) is plotted {\em vs.}\/ rescaled time $t^{d/\alpha}$. With this rescaling, the logarithm of the bound in \eqref{e:upperbound} is a straight line (upper red line in the right plot). $Q_\Lambda$, evaluated for chains of lengths $N=25001$, 50001, and 100001 are shown in blue, but the curves fall onto each other and are indistinguishable on the scale of the plot. The linearly decaying trend in the plot, superimposed by fluctuations, confirms over a range of more than hundred orders of magnitude that the bound in \eqref{e:upperbound} correctly captures the compressed or stretched exponential decay of the long-time asymptotics of $P_\Lambda$. Again, the numerical constants in the compressed or stretched exponential decay are, as expected, overestimated.
}%
\end{figure}

For any given $t$, the inequality
\begin{equation}
\frac{2\lvert Jt\rvert }{\lvert k\rvert ^\alpha}<\frac{\pi}{2}\qquad\forall k\notin B_{R(t)}
\end{equation}
holds, where
\begin{equation}\label{e:Rt}
R(t)=\left\lvert\frac{4Jt}{\pi}\right\rvert^{1/\alpha}
\end{equation}
and
\begin{equation}
B_{R} = \left\{k\in\Lambda\,\big|\,\lvert k\rvert <R\right\}
\end{equation}
is a $d$-dimensional ball of radius $R$, centered around the origin. Hence we can split the product \eqref{e:P_N} in the following way,
\begin{align}\label{e:P3}
P_\Lambda&= \prod_{k\in B_{R(t)}\setminus0} \underbrace{\left\lvert \cos\left(\frac{2Jt}{\lvert k\rvert ^\alpha}\right)\right\rvert }_{\displaystyle\leq1}
\prod_{k\in\Lambda\setminus B_{R(t)}} \left\lvert \cos\left(\frac{2Jt}{\lvert k\rvert ^\alpha}\right)\right\rvert \nonumber\\
&\leq\prod_{k\in\Lambda\setminus B_{R(t)}} \left\lvert \cos\left(\frac{2Jt}{\lvert k\rvert ^\alpha}\right)\right\rvert .
\end{align}
For finite lattices and some large $t$, it will happen that the remaining product in \eqref{e:P3} consists of no factors at all, resulting in a trivial upper bound $P_\Lambda\leq1$. In fact, this was to be expected, since the unitary, quasiperiodic dynamics of finite systems gives rise to recurrences. But we are interested in the thermodynamic limit where, for a given $t$ and large enough lattice $\Lambda$, the remaining product in \eqref{e:P3} will consist of a large (infinite) number of factors. By virtue of the definition of $R(t)$ in \eqref{e:Rt}, the arguments of the cosines in these factors are all between $-\pi/2$ and $\pi/2$, and for this range of arguments the cosine can be bounded by
\begin{equation}
\lvert \cos(x)\rvert \leq1-\frac{4x^2}{\pi^2}.
\end{equation}
From elementary properties of the exponential function it then follows that, in the limit of large lattice size,
\begin{align}\label{e:expsum}
P_\Lambda&\leq\prod_{k\in\Lambda\setminus B_{R(t)}} \left[1-\left(\frac{4Jt}{\pi \lvert k\rvert^\alpha}\right)^2\right]\nonumber\\
&\leq\prod_{k\in\Lambda\setminus B_{R(t)}} \exp\left[-\left(\frac{4Jt}{\pi \lvert k\rvert^\alpha}\right)^2\right]\nonumber\\
&= \exp\Biggl[-\left(\frac{4Jt}{\pi}\right)^2\sum_{k\in\Lambda\setminus B_{R(t)}} \frac{1}{\lvert k\rvert^{2\alpha}}\Biggr].
\end{align}
Asymptotically for large lattices of ``radius'' $\mathcal{R}=N^{1/d}/2$, the sum in \eqref{e:expsum} can be determined by integral approximation
\begin{multline}\label{e:sumapprox}
\sum_{k\in\Lambda\setminus B_{R(t)}} \frac{1}{\lvert k\rvert^{2\alpha}}\sim\int_{B_\mathcal{R}\setminus B_{R(t)}}\rmd^d r\, r^{-2\alpha}\\
\geq \frac{2\pi^{d/2}}{\Gamma(d/2)}\int_{R(t)+1}^\mathcal{R}\rmd r\, r^{d-2\alpha-1}\\
= \frac{2\pi^{d/2}}{\Gamma(d/2)}\frac{2^{2\alpha-d}N^{1-2\alpha/d}-(\lvert 4Jt/\pi\rvert^{1/\alpha}+1)^{d-2\alpha}}{d-2\alpha}\\
> \frac{2\pi^{d/2}}{\Gamma(d/2)}\frac{2^{2\alpha-d}N^{1-2\alpha/d}-2^{d-2\alpha}\lvert 4Jt/\pi\rvert^{d/\alpha-2}}{d-2\alpha}\\
\sim\frac{2\pi^{d/2}}{\vert d-2\alpha\rvert\Gamma(d/2)}\begin{cases}
2^{2\alpha-d}N^{1-2\alpha/d} & \text{for $\alpha<d/2$},\\
2^{d-2\alpha}\lvert 4Jt/\pi\rvert^{d/\alpha-2} & \text{for $\alpha>d/2$},
\end{cases}
\end{multline}
where in the last line only the leading contribution in the limit of large lattice size $N$ was retained. For the second inequality in \eqref{e:sumapprox} we have used that
\begin{equation}
\frac{\left(x^{1/\alpha}+1\right)^{d-2\alpha}}{2\alpha-d}>\frac{2^{d-2\alpha}x^{d/\alpha-2}}{2\alpha-d}\qquad\text{for $x>1$}.
\end{equation}
This bound can easily be tightened, at the cost of pushing its validity out to larger values of $x$. Inserting \eqref{e:sumapprox} into \eqref{e:expsum} yields the bound
\begin{equation}\label{e:upperbound}
P_\Lambda\leq
\begin{cases}
\exp\left[-\frac{2^{5+2\alpha-d}\pi^{d/2-2}J^2}{(d-2\alpha)\Gamma(d/2)}t^2 N^{1-2\alpha/d}\right] & \text{for $\alpha<d/2$},\\[2mm]
\exp\left[-\frac{2^{1+d-2\alpha}\pi^{d/2}}{(2\alpha-d)\Gamma(d/2)}\left\lvert\frac{4Jt}{\pi}\right\rvert^{d/\alpha}\right] & \text{for $\alpha>d/2$},
\end{cases}
\end{equation}
valid for $4\lvert Jt\rvert>\pi$ and in the limit of large lattice size $N$. Inserting this bound into \eqref{e:xfinal}, we obtain the bound on the modulus of the expectation value $\left\langle\sigma_i^x\right\rangle$ as given in Eq.\ (3) of the main paper. A comparison of these bounds with an exact evaluation of \eqref{e:P_N} for finite lattices is shown, over more than hundred orders of magnitude, in Fig.\ \ref{f:bounds}.

\section{C.\hspace{2mm} More simulation results for the \texorpdfstring{$\alpha XY$}{alphaXY} chain}

In the main paper, relaxation time scales of the $\alpha XY$ model have been analyzed for $e=25/96\approx0.26$, an energy density in the unmagnetized phase, located slightly above the phase transition energy $e_\text{c}=1/4$. In the present section we will report simulation results also for other values of $e$, both in the magnetized and in the unmagnetized phase. The simulation methods used are the same as described in the main paper.

\subsection{C.1\hspace{2mm} Magnetized phase}

\begin{figure}\centering
\includegraphics[width=0.8\linewidth]{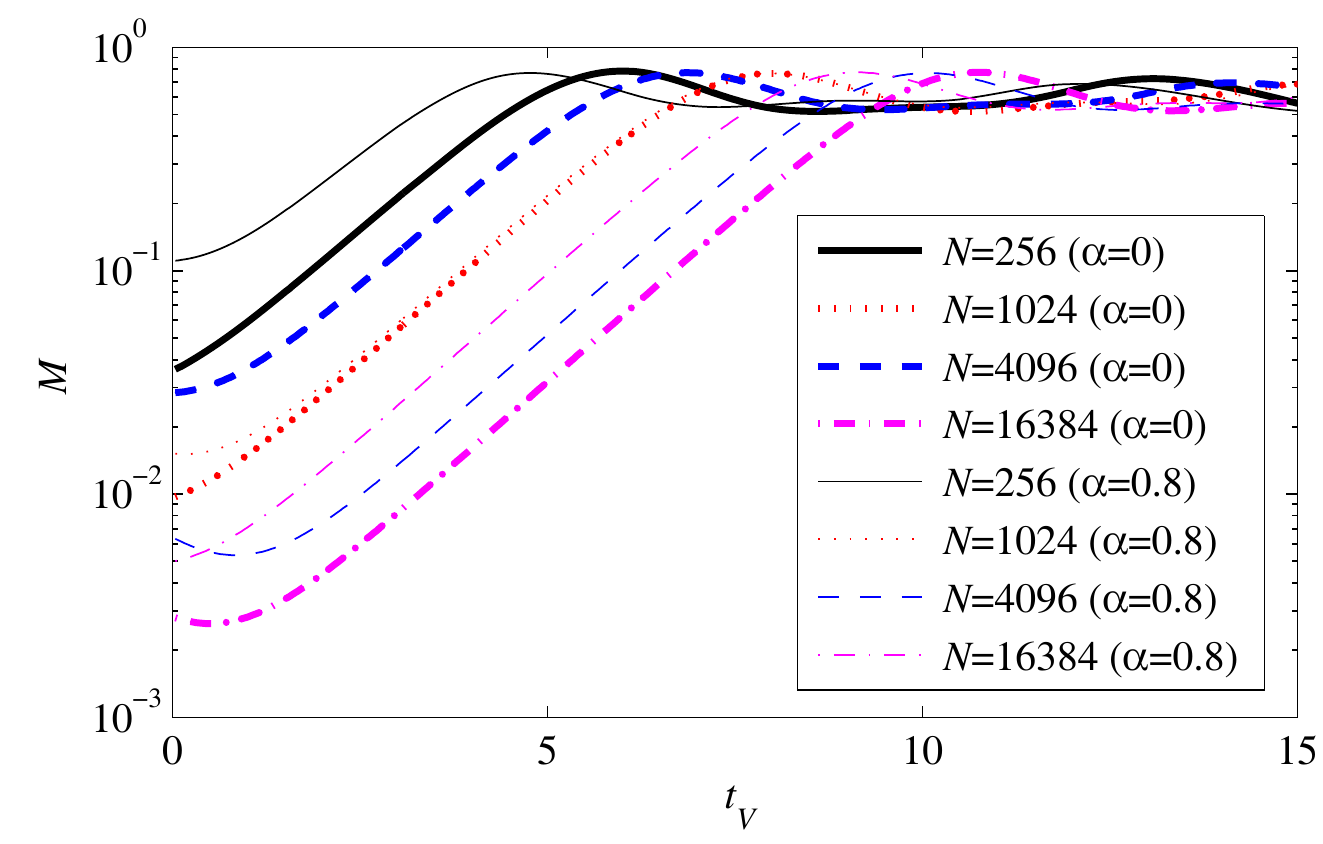}
\caption{\label{fig:transient}
(Color online) The magnetization of the $\alpha XY$ chain, plotted for $\alpha=0.0$ (thick lines) and  $\alpha=0.8$ (thin lines), for $\Delta=1/5$, and system sizes $N=256$, $N=1024$, $4096$ and $16384$. $t_V$ refers to the time of the Vlasov approximation, i.e., of the normalized Hamiltonian with the factor $\mathcal{N}$.
}
\end{figure}

For energy densities $e\approx0.007$ well below $e_\text{c}=1/4$, and starting from unmagnetized out-of-equilibrium initial conditions, we observe in our simulations that the system experiences a fast transient to a magnetized state, and then remains magnetized at all times. This transient corresponds to a regime where the macroscopic evolution of the system is well captured by a Vlasov equation. The initial growth of the magnetization corresponds to a linear instability of the Vlasov equation~\cite{Yamaguchi_etal04}, and the exponential rate neither depends significantly on the system size, nor on the interaction range $\alpha$~\cite{Bachelard2011} (see Fig.~\ref{fig:transient}). Consequently the time needed by the magnetization to reach a finite value is determined mainly by the initial condition, in particular the initial value of the magnetization which typically scales as $1/\sqrt{N}$. We point out that this cannot account for the threshold discussed in the main paper. Firstly because, in the mean-field case, the power laws of the relaxation times were shown to originate from kinetic terms that are not present in the Vlasov equation~\cite{BouchetDauxois05}. Secondly, the Vlasov equation holds for the Hamiltonian normalized with the factor $\mathcal{N}$, so there is actually a trivial $N$-dependence $q=(\alpha-1)/2$ for the exponential rate, plus the $1/2$ from the  $1/\sqrt{N}$ of the fluctuations; nevertheless, these factors do not yield any threshold at $\alpha=d/2$.

\begin{figure}\centering
\includegraphics[width=0.48\linewidth]{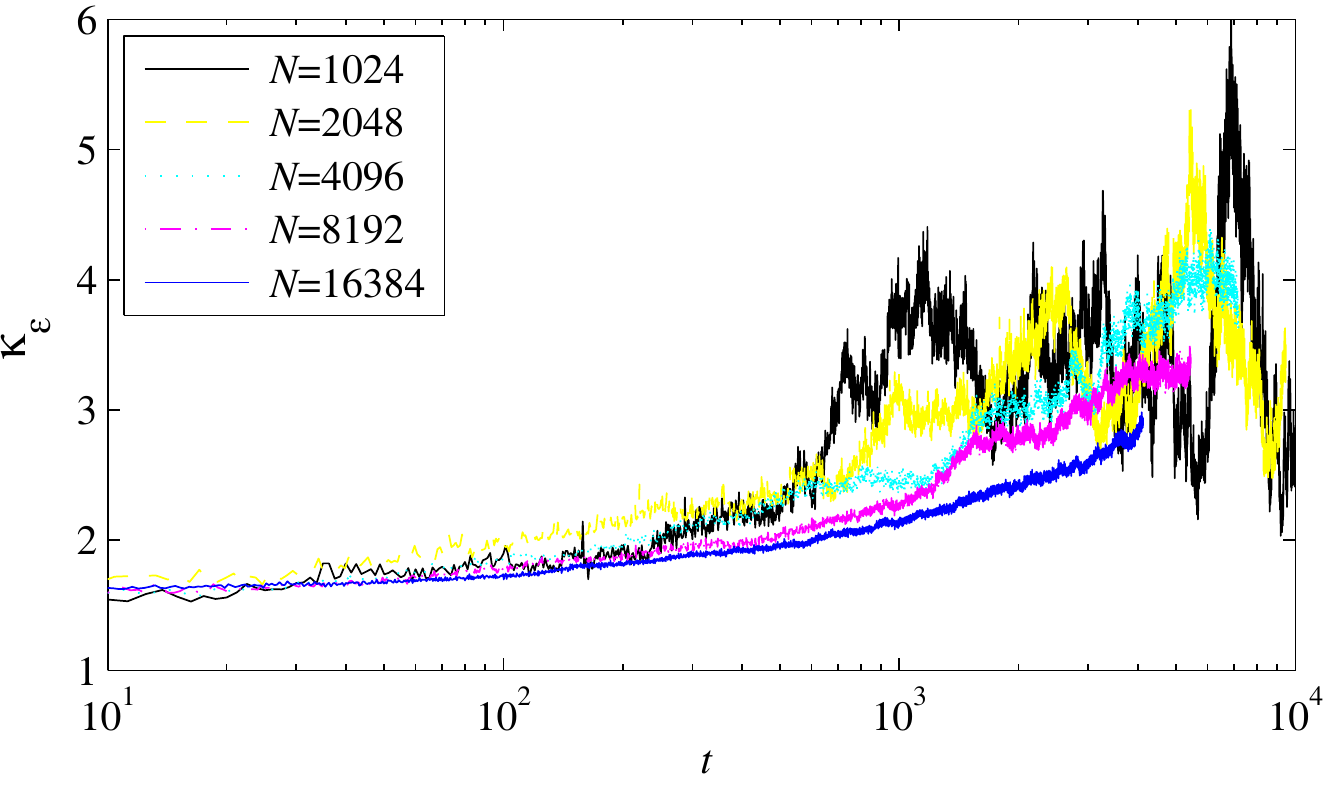}
\includegraphics[width=0.48\linewidth]{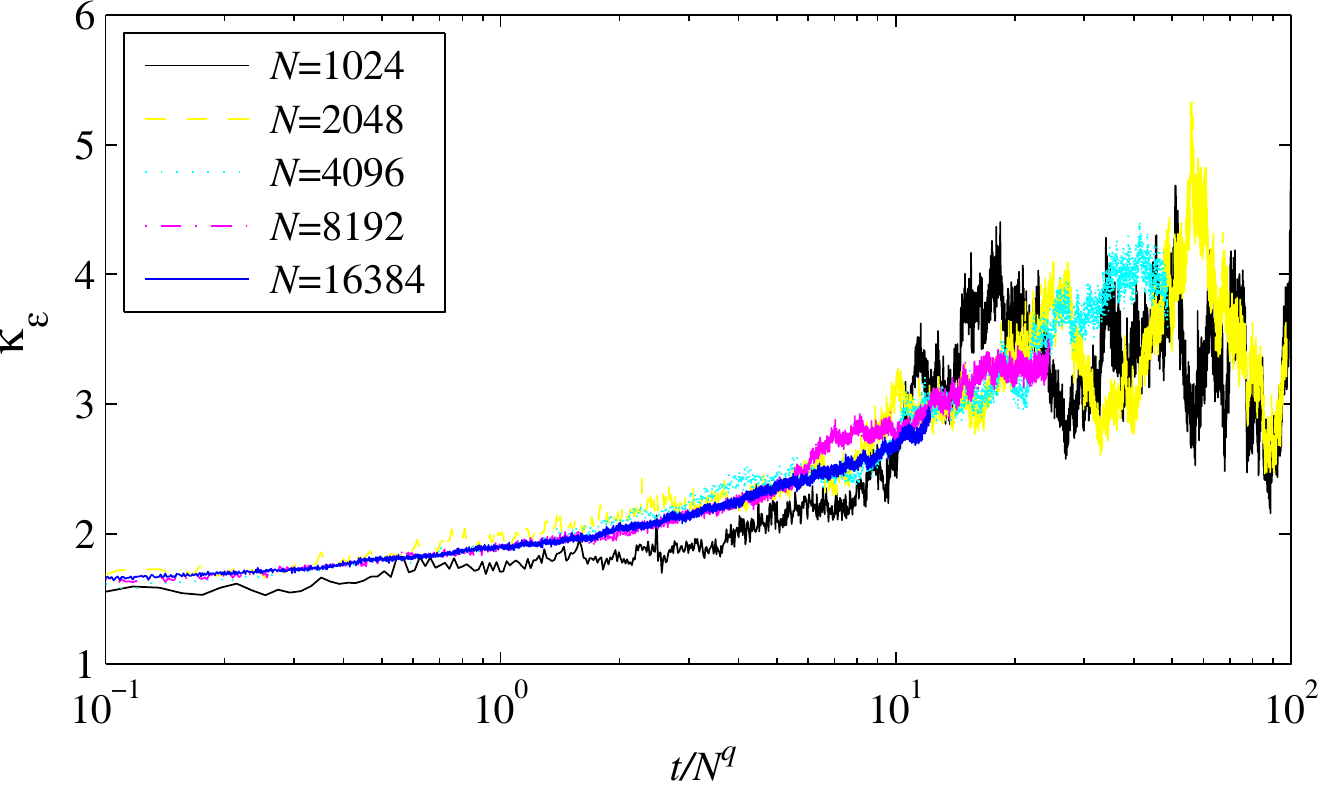}
\caption{\label{fig:unmag}
(Color online) The kurtosis $\kappa_\epsilon$ of the $\epsilon$-distribution of the $\alpha XY$ chain, plotted for $\alpha=0.2$ and $\Delta=1/5$, i.e. energy density $e\approx0.007$, and system sizes $N=1024$, $2048$, $4096$, $8192$ and $16384$. Data are averaged over $5$ realizations of waterbag initial conditions. Left: As a function of time $t$, the time scale on which $\kappa_\epsilon$ relaxes to its equilibrium value grows with the system size $N$. Right: As before, but as a function of rescaled time $t/N^q$ with scaling exponent $q=0.6$.
}
\end{figure}

Apart from the fast transient, the magnetization, temperature, and kurtosis of the momentum distribution all change only very slightly during the relaxation process. As a result, their relaxation is virtually impossible to observe in the presence of finite-size fluctuations, and we need to look for a more suitable observable. To this end we introduce the normalized single particle energy at the lattice site $i$,
\begin{equation}
\epsilon_i=\frac{p_i^2}{2}-\frac{J}{N^{1-\alpha}}\sum_{\substack{j=1\\j\neq i}}^N\frac{\cos(\phi_i-\phi_j)}{|i-j|^\alpha},
\end{equation}
whose kurtosis $\kappa_\epsilon$ exhibits a slow relaxation over a large range of values (see Fig.~\ref{fig:unmag}), which is good basis for observing the relaxation dynamics in the presence of fluctuations. Unfortunately, as can be seen in that same figure, the fluctuations in $\kappa_\epsilon$ are also fairly large, and this prevents a good data collapse when rescaling time as $tN^{-q(\alpha)}$. This suggests that much larger systems need to be studied in order to extract scaling laws for the equilibration times in this regime to good accuracy, and this will require significantly more computational power.
\begin{figure}[b]\centering
\includegraphics[width=0.48\linewidth]{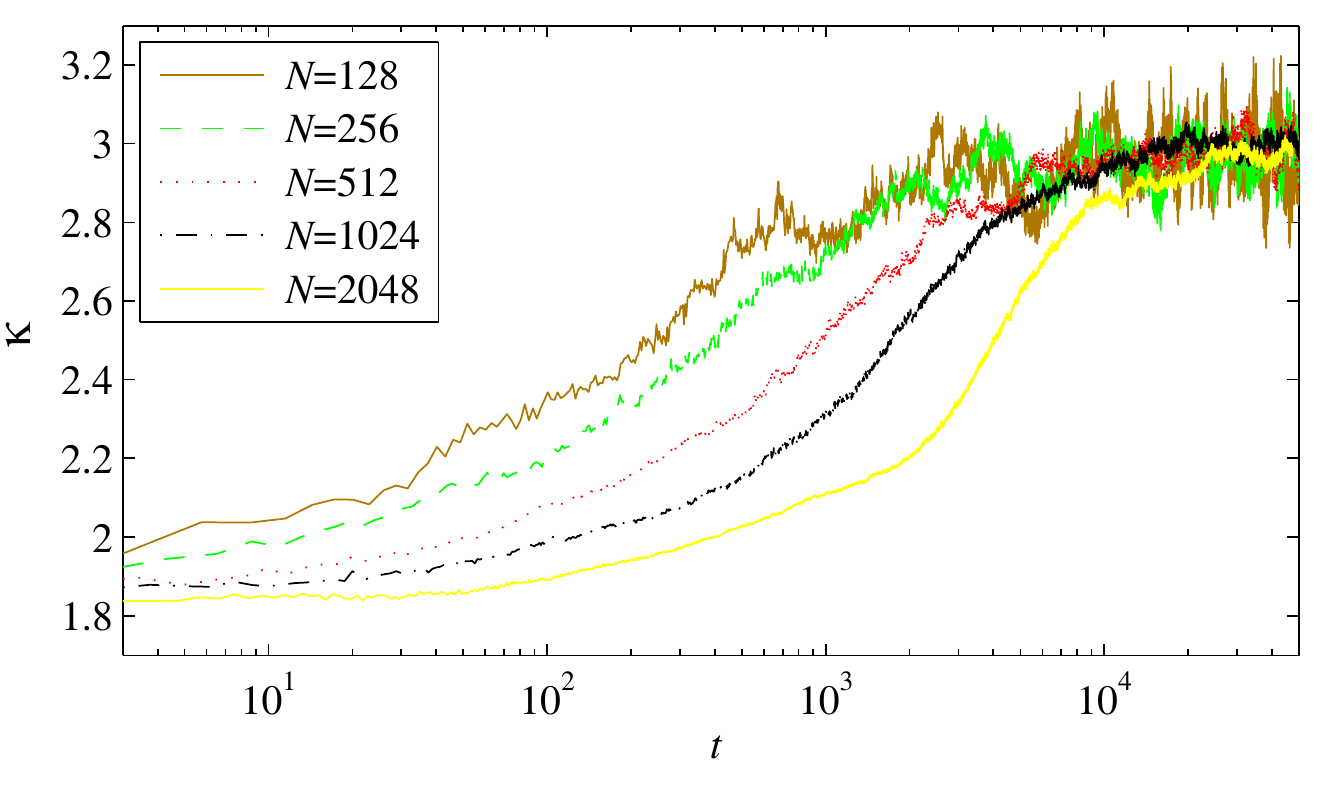}
\includegraphics[width=0.48\linewidth]{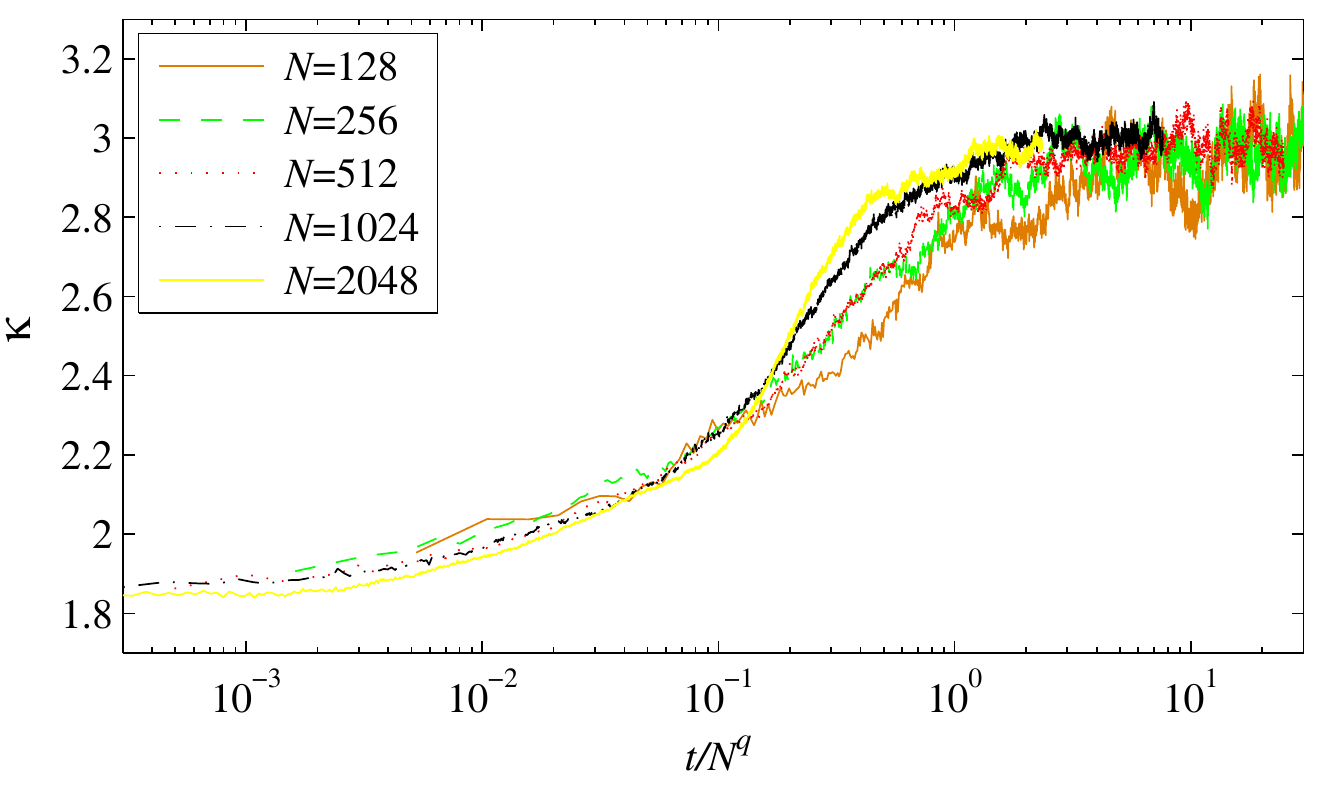}
\caption{\label{fig:transition}
(Color online) The kurtosis $\kappa$ of the momentum distribution of the $\alpha XY$ chain, plotted for $\alpha=0.2$ and $\Delta=1.03$, i.e. energy density $e\approx0.18$, and system sizes $N=128$, $N=256$, $N=512$, $N=1024$ and $2048$. Data are averaged over $10$ realizations of waterbag initial conditions. Left: As a function of time $t$, the time scale on which $\kappa_\epsilon$ relaxes to its equilibrium value grows with the system size $N$. Right: As before, but as a function of rescaled time $t/N^q$ with scaling exponent $q=1.3$.
}
\end{figure}

Next we study the relaxation time scales at an energy density $e\approx0.18$ still in the magnetized phase, but only slightly below the critical energy. At this energy we observe that the system remains unmagnetized for a long time before finally switching to the nonzero equilibrium value of the magnetization in the magnetized phase (see Fig.~\ref{fig:transition}). In this regime, the $N$-dependent time scales of the Hamiltonian Mean-Field model ($\alpha XY$ model with $\alpha=0$) were originally studied in \cite{Yamaguchi03} and the exponent $q=1.2$ was extracted by means of a four-parameter fit to the magnetization. The result of such a fit depends considerably on the choice of the fitting function.

\begin{figure}\centering
\includegraphics[width=0.9\linewidth]{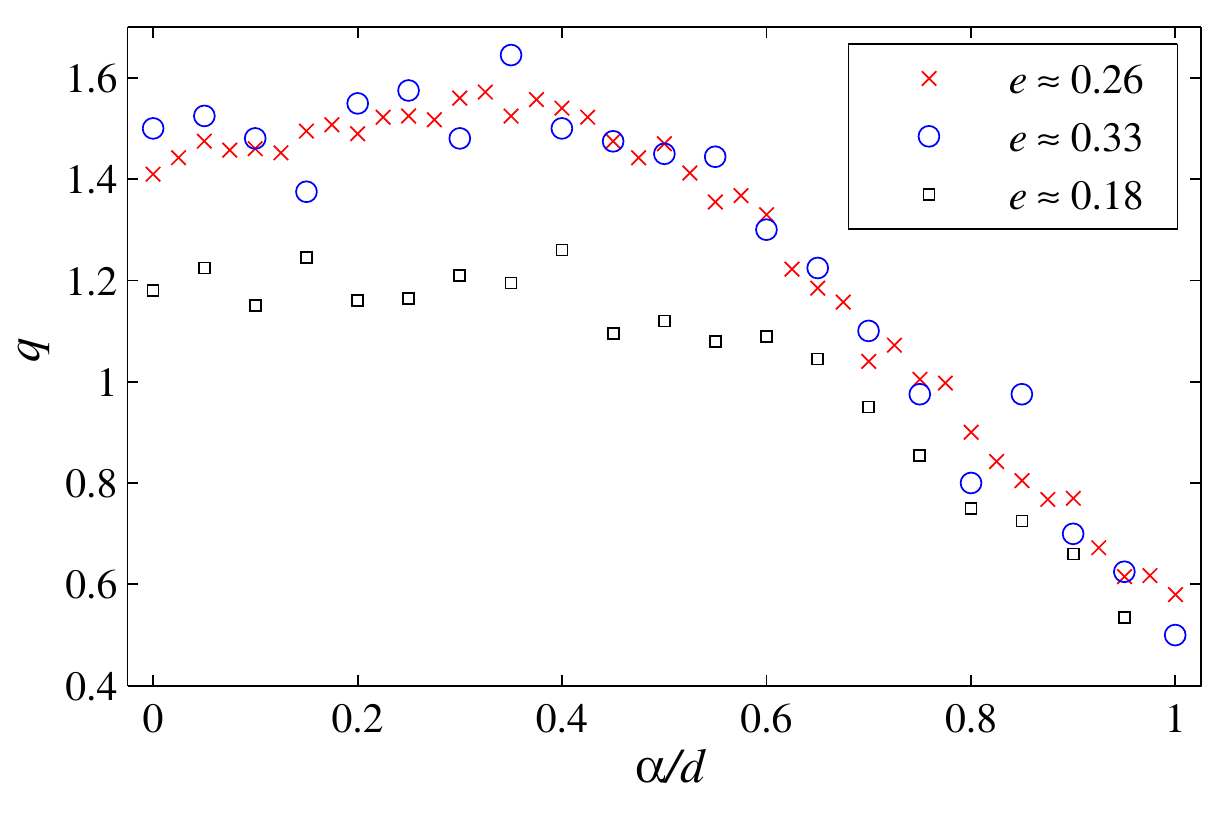}
\caption{\label{fig:scalinglaws}
(Color online) The exponent $q$ of the scaling law $\tau\propto N^q$ that governs the system-size dependence of the relaxation time $\tau$ in the $\alpha XY$ long-range chain. The red crosses correspond to the unmagnetized regime $e\approx0.26$, the blue circle to the higher energy regime $e\approx0.33$, the black squares to the magnetized regime, for $e\approx0.007$. Data are averaged over $10$ realizations for the $e\approx0.33$ case, and over $5$ realizations in the $e\approx0.007$ case.
}
\end{figure}

To avoid the arbitrariness of a fitting function, we use a simple rescaling of time as for the other energies studied, although it leads to a weaker collapse (see Fig.~\ref{fig:transition}). Still, with some optimism, the exponent $q$ of the scaling law $\tau\propto N^q$ can be extracted from the data (see Fig.\ \ref{fig:scalinglaws}). Despite the large fluctuations in $q$, the change of regime at around $\alpha=d/2$ is at least vaguely perceptible and not inconsistent with the universal threshold we propose. The exponent $q=1.2$ obtained in \cite{Yamaguchi03} for the magnetized regime is smaller than the value $q=1.41$ we obtain for $\alpha=0$ in the unmagnetized regime, but both findings are consistent with the results of Bouchet and Dauxois \cite{BouchetDauxois05} where---translated into our notation---it is proved analytically that $q$ is larger than $1/2$ in this regime.

\subsection{C.2\hspace{2mm} Unmagnetized phase}

In the unmagnetized phase above the critical energy, we choose the kurtosis $\kappa$ of the momentum distribution as an observable, as described in the main paper. In addition to the energy value $e=25/96$ slightly above the phase transition energy $e_\text{c}=1/4$, we performed simulations also at higher energy densities, and we observe the following antagonistic effects: The higher the energy, the smaller are the fluctuations of $\kappa$ in the simulations, and the better is the data collapse when rescaling time. However, at these higher energies longer simulation times are necessary to reach equilibrium. These longer relaxation times could be avoided by considering smaller system sizes, but this in turn would increase the fluctuations. As a consequence, a balance must be found between simulation times that can realistically be reached on the one hand, and accessible system sizes and/or energies on the other hand. The choice $e\approx0.26$ as reported in the main paper seemed a good compromise and yields a very good data collapse of $\kappa$ and small errorbars for the scaling exponent $q$. In Fig.~\ref{fig:scalinglaws} we show results for $q(\alpha)$ at a higher energy density $e\approx0.33$ ($\Delta=1.4$). The results are similar to the $e\approx0.26$ case, but with larger fluctuations.

In summary, for all the energy densities we have sampled, the simulation results are qualitatively similar to the $e\approx0.26$ case discussed in the main paper. The results are consistent with the presence of a universal threshold value at $\alpha_\text{th}=1/2$ as suggested in the main paper, but large fluctuations prevent us from drawing stronger conclusions.

\bibliography{DLRThreshold}

\begin{thebibliography}{32}%
\makeatletter
\providecommand \@ifxundefined [1]{%
 \@ifx{#1\undefined}
}%
\providecommand \@ifnum [1]{%
 \ifnum #1\expandafter \@firstoftwo
 \else \expandafter \@secondoftwo
 \fi
}%
\providecommand \@ifx [1]{%
 \ifx #1\expandafter \@firstoftwo
 \else \expandafter \@secondoftwo
 \fi
}%
\providecommand \natexlab [1]{#1}%
\providecommand \enquote  [1]{``#1''}%
\providecommand \bibnamefont  [1]{#1}%
\providecommand \bibfnamefont [1]{#1}%
\providecommand \citenamefont [1]{#1}%
\providecommand \href@noop [0]{\@secondoftwo}%
\providecommand \href [0]{\begingroup \@sanitize@url \@href}%
\providecommand \@href[1]{\@@startlink{#1}\@@href}%
\providecommand \@@href[1]{\endgroup#1\@@endlink}%
\providecommand \@sanitize@url [0]{\catcode `\\12\catcode `\$12\catcode
  `\&12\catcode `\#12\catcode `\^12\catcode `\_12\catcode `\%12\relax}%
\providecommand \@@startlink[1]{}%
\providecommand \@@endlink[0]{}%
\providecommand \url  [0]{\begingroup\@sanitize@url \@url }%
\providecommand \@url [1]{\endgroup\@href {#1}{\urlprefix }}%
\providecommand \urlprefix  [0]{URL }%
\providecommand \Eprint [0]{\href }%
\providecommand \doibase [0]{http://dx.doi.org/}%
\providecommand \selectlanguage [0]{\@gobble}%
\providecommand \bibinfo  [0]{\@secondoftwo}%
\providecommand \bibfield  [0]{\@secondoftwo}%
\providecommand \translation [1]{[#1]}%
\providecommand \BibitemOpen [0]{}%
\providecommand \bibitemStop [0]{}%
\providecommand \bibitemNoStop [0]{.\EOS\space}%
\providecommand \EOS [0]{\spacefactor3000\relax}%
\providecommand \BibitemShut  [1]{\csname bibitem#1\endcsname}%
\let\auto@bib@innerbib\@empty
\bibitem [{\citenamefont {Lynden-Bell}\ and\ \citenamefont
  {Wood}(1968)}]{LynWood68}%
  \BibitemOpen
  \bibfield  {author} {\bibinfo {author} {\bibfnamefont {D.}~\bibnamefont
  {Lynden-Bell}}\ and\ \bibinfo {author} {\bibfnamefont {R.}~\bibnamefont
  {Wood}},\ }\href@noop {} {\bibfield  {journal} {\bibinfo  {journal} {Mon.
  Not. R. Astron. Soc.}\ }\textbf {\bibinfo {volume} {138}},\ \bibinfo {pages}
  {495} (\bibinfo {year} {1968})}\BibitemShut {NoStop}%
\bibitem [{\citenamefont {Thirring}(1970)}]{Thirring70}%
  \BibitemOpen
  \bibfield  {author} {\bibinfo {author} {\bibfnamefont {W.}~\bibnamefont
  {Thirring}},\ }\href {\doibase 10.1007/BF01403177} {\bibfield  {journal}
  {\bibinfo  {journal} {Z. Phys.}\ }\textbf {\bibinfo {volume} {235}},\
  \bibinfo {pages} {339} (\bibinfo {year} {1970})}\BibitemShut {NoStop}%
\bibitem [{\citenamefont {Campa}\ \emph {et~al.}(2009)\citenamefont {Campa},
  \citenamefont {Dauxois},\ and\ \citenamefont {Ruffo}}]{CamDauxRuf09}%
  \BibitemOpen
  \bibfield  {author} {\bibinfo {author} {\bibfnamefont {A.}~\bibnamefont
  {Campa}}, \bibinfo {author} {\bibfnamefont {T.}~\bibnamefont {Dauxois}}, \
  and\ \bibinfo {author} {\bibfnamefont {S.}~\bibnamefont {Ruffo}},\ }\href
  {\doibase 10.1016/j.physrep.2009.07.001} {\bibfield  {journal} {\bibinfo
  {journal} {Phys. Rep.}\ }\textbf {\bibinfo {volume} {480}},\ \bibinfo {pages}
  {57} (\bibinfo {year} {2009})}\BibitemShut {NoStop}%
\bibitem [{\citenamefont {Ruelle}(1969)}]{Ruelle}%
  \BibitemOpen
  \bibfield  {author} {\bibinfo {author} {\bibfnamefont {D.}~\bibnamefont
  {Ruelle}},\ }\href@noop {} {\emph {\bibinfo {title} {Statistical Mechanics:
  {R}igorous Results}}}\ (\bibinfo  {publisher} {Benjamin, Reading},\ \bibinfo
  {year} {1969})\BibitemShut {NoStop}%
\bibitem [{\citenamefont {Antoni}\ and\ \citenamefont {Ruffo}(1995)}]{AnRu95}%
  \BibitemOpen
  \bibfield  {author} {\bibinfo {author} {\bibfnamefont {M.}~\bibnamefont
  {Antoni}}\ and\ \bibinfo {author} {\bibfnamefont {S.}~\bibnamefont {Ruffo}},\
  }\href {\doibase 10.1103/PhysRevE.52.2361} {\bibfield  {journal} {\bibinfo
  {journal} {Phys. Rev. E}\ }\textbf {\bibinfo {volume} {52}},\ \bibinfo
  {pages} {2361} (\bibinfo {year} {1995})}\BibitemShut {NoStop}%
\bibitem [{\citenamefont {Joyce}\ and\ \citenamefont
  {Worrakitpoonpon}(2010)}]{JoyceWorrakitpoonpon10}%
  \BibitemOpen
  \bibfield  {author} {\bibinfo {author} {\bibfnamefont {M.}~\bibnamefont
  {Joyce}}\ and\ \bibinfo {author} {\bibfnamefont {T.}~\bibnamefont
  {Worrakitpoonpon}},\ }\href {\doibase 10.1088/1742-5468/2010/10/P10012}
  {\bibfield  {journal} {\bibinfo  {journal} {J. Stat. Mech.}\ }\textbf
  {\bibinfo {volume} {2010}},\ \bibinfo {pages} {P10012} (\bibinfo {year}
  {2010})}\BibitemShut {NoStop}%
\bibitem [{\citenamefont {O'Dell}\ \emph {et~al.}(2000)\citenamefont {O'Dell},
  \citenamefont {Giovanazzi}, \citenamefont {Kurizki},\ and\ \citenamefont
  {Akulin}}]{ODell_etal00}%
  \BibitemOpen
  \bibfield  {author} {\bibinfo {author} {\bibfnamefont {D.}~\bibnamefont
  {O'Dell}}, \bibinfo {author} {\bibfnamefont {S.}~\bibnamefont {Giovanazzi}},
  \bibinfo {author} {\bibfnamefont {G.}~\bibnamefont {Kurizki}}, \ and\
  \bibinfo {author} {\bibfnamefont {V.~M.}\ \bibnamefont {Akulin}},\ }\href
  {\doibase 10.1103/PhysRevLett.84.5687} {\bibfield  {journal} {\bibinfo
  {journal} {Phys. Rev. Lett.}\ }\textbf {\bibinfo {volume} {84}},\ \bibinfo
  {pages} {5687} (\bibinfo {year} {2000})}\BibitemShut {NoStop}%
\bibitem [{\citenamefont {Dominguez}\ \emph {et~al.}(2010)\citenamefont
  {Dominguez}, \citenamefont {Oettel},\ and\ \citenamefont
  {Dietrich}}]{Dominguez_etal10}%
  \BibitemOpen
  \bibfield  {author} {\bibinfo {author} {\bibfnamefont {A.}~\bibnamefont
  {Dominguez}}, \bibinfo {author} {\bibfnamefont {M.}~\bibnamefont {Oettel}}, \
  and\ \bibinfo {author} {\bibfnamefont {S.}~\bibnamefont {Dietrich}},\ }\href
  {\doibase 10.1103/PhysRevE.82.011402} {\bibfield  {journal} {\bibinfo
  {journal} {Phys. Rev. E}\ }\textbf {\bibinfo {volume} {82}},\ \bibinfo
  {pages} {11402} (\bibinfo {year} {2010})}\BibitemShut {NoStop}%
\bibitem [{\citenamefont {Golestanian}(2012)}]{Golestanian12}%
  \BibitemOpen
  \bibfield  {author} {\bibinfo {author} {\bibfnamefont {R.}~\bibnamefont
  {Golestanian}},\ }\href {\doibase 10.1103/PhysRevLett.108.038303} {\bibfield
  {journal} {\bibinfo  {journal} {Phys. Rev. Lett.}\ }\textbf {\bibinfo
  {volume} {108}},\ \bibinfo {pages} {038303} (\bibinfo {year}
  {2012})}\BibitemShut {NoStop}%
\bibitem [{\citenamefont {Chalony}\ \emph {et~al.}(2013)\citenamefont
  {Chalony}, \citenamefont {Barr\'e}, \citenamefont {Marcos}, \citenamefont
  {Olivetti},\ and\ \citenamefont {Wilkowski}}]{Chalony_etal12}%
  \BibitemOpen
  \bibfield  {author} {\bibinfo {author} {\bibfnamefont {M.}~\bibnamefont
  {Chalony}}, \bibinfo {author} {\bibfnamefont {J.}~\bibnamefont {Barr\'e}},
  \bibinfo {author} {\bibfnamefont {B.}~\bibnamefont {Marcos}}, \bibinfo
  {author} {\bibfnamefont {A.}~\bibnamefont {Olivetti}}, \ and\ \bibinfo
  {author} {\bibfnamefont {D.}~\bibnamefont {Wilkowski}},\ }\href {\doibase
  10.1103/PhysRevA.87.013401} {\bibfield  {journal} {\bibinfo  {journal} {Phys.
  Rev. A}\ }\textbf {\bibinfo {volume} {87}},\ \bibinfo {pages} {013401}
  (\bibinfo {year} {2013})}\BibitemShut {NoStop}%
\bibitem [{\citenamefont {Britton}\ \emph {et~al.}(2012)\citenamefont
  {Britton}, \citenamefont {Sawyer}, \citenamefont {Keith}, \citenamefont
  {Wang}, \citenamefont {Freericks}, \citenamefont {Uys}, \citenamefont
  {Biercuk},\ and\ \citenamefont {Bollinger}}]{Britton_etal12}%
  \BibitemOpen
  \bibfield  {author} {\bibinfo {author} {\bibfnamefont {J.~W.}\ \bibnamefont
  {Britton}}, \bibinfo {author} {\bibfnamefont {B.~C.}\ \bibnamefont {Sawyer}},
  \bibinfo {author} {\bibfnamefont {A.~C.}\ \bibnamefont {Keith}}, \bibinfo
  {author} {\bibfnamefont {C.-C.~J.}\ \bibnamefont {Wang}}, \bibinfo {author}
  {\bibfnamefont {J.~K.}\ \bibnamefont {Freericks}}, \bibinfo {author}
  {\bibfnamefont {H.}~\bibnamefont {Uys}}, \bibinfo {author} {\bibfnamefont
  {M.~J.}\ \bibnamefont {Biercuk}}, \ and\ \bibinfo {author} {\bibfnamefont
  {J.~J.}\ \bibnamefont {Bollinger}},\ }\href {\doibase 10.1038/nature10981}
  {\bibfield  {journal} {\bibinfo  {journal} {Nature}\ }\textbf {\bibinfo
  {volume} {484}},\ \bibinfo {pages} {489} (\bibinfo {year}
  {2012})}\BibitemShut {NoStop}%
\bibitem [{\citenamefont {Kastner}(2011)}]{Kastner11}%
  \BibitemOpen
  \bibfield  {author} {\bibinfo {author} {\bibfnamefont {M.}~\bibnamefont
  {Kastner}},\ }\href {\doibase 10.1103/PhysRevLett.106.130601} {\bibfield
  {journal} {\bibinfo  {journal} {Phys. Rev. Lett}\ }\textbf {\bibinfo {volume}
  {106}},\ \bibinfo {pages} {130601} (\bibinfo {year} {2011})}\BibitemShut
  {NoStop}%
\bibitem [{\citenamefont {Kastner}(2012)}]{Kastner12}%
  \BibitemOpen
  \bibfield  {author} {\bibinfo {author} {\bibfnamefont {M.}~\bibnamefont
  {Kastner}},\ }\href {\doibase 10.2478/s11534-011-0122-4} {\bibfield
  {journal} {\bibinfo  {journal} {Cent. Eur. J. Phys.}\ }\textbf {\bibinfo
  {volume} {10}},\ \bibinfo {pages} {637} (\bibinfo {year} {2012})}\BibitemShut
  {NoStop}%
\bibitem [{\citenamefont {Emch}(1966)}]{Emch66}%
  \BibitemOpen
  \bibfield  {author} {\bibinfo {author} {\bibfnamefont {G.~G.}\ \bibnamefont
  {Emch}},\ }\href {\doibase 10.1063/1.1705023} {\bibfield  {journal} {\bibinfo
   {journal} {J. Math. Phys.}\ }\textbf {\bibinfo {volume} {7}},\ \bibinfo
  {pages} {1198} (\bibinfo {year} {1966})}\BibitemShut {NoStop}%
\bibitem [{\citenamefont {Radin}(1970)}]{Radin70}%
  \BibitemOpen
  \bibfield  {author} {\bibinfo {author} {\bibfnamefont {C.}~\bibnamefont
  {Radin}},\ }\href {\doibase 10.1063/1.1665079} {\bibfield  {journal}
  {\bibinfo  {journal} {J. Math. Phys.}\ }\textbf {\bibinfo {volume} {11}},\
  \bibinfo {pages} {2945} (\bibinfo {year} {1970})}\BibitemShut {NoStop}%
\bibitem [{\citenamefont {van~den Worm}\ \emph {et~al.}()\citenamefont {van~den
  Worm}, \citenamefont {Sawyer}, \citenamefont {Bollinger},\ and\ \citenamefont
  {Kastner}}]{vandenWorm_etal}%
  \BibitemOpen
  \bibfield  {author} {\bibinfo {author} {\bibfnamefont {M.}~\bibnamefont
  {van~den Worm}}, \bibinfo {author} {\bibfnamefont {B.~C.}\ \bibnamefont
  {Sawyer}}, \bibinfo {author} {\bibfnamefont {J.~J.}\ \bibnamefont
  {Bollinger}}, \ and\ \bibinfo {author} {\bibfnamefont {M.}~\bibnamefont
  {Kastner}},\ }\href@noop {} {\enquote {\bibinfo {title} {Relaxation
  timescales and decay of correlations in a long-range interacting quantum
  simulator},}\ }\Eprint {http://arxiv.org/abs/1209.3697} {arXiv:1209.3697}
  \BibitemShut {NoStop}%
\bibitem [{\citenamefont {Anteneodo}\ and\ \citenamefont
  {Tsallis}(1998)}]{AnteneodoTsallis98}%
  \BibitemOpen
  \bibfield  {author} {\bibinfo {author} {\bibfnamefont {C.}~\bibnamefont
  {Anteneodo}}\ and\ \bibinfo {author} {\bibfnamefont {C.}~\bibnamefont
  {Tsallis}},\ }\href {\doibase 10.1103/PhysRevLett.80.5313} {\bibfield
  {journal} {\bibinfo  {journal} {Phys. Rev. Lett.}\ }\textbf {\bibinfo
  {volume} {80}},\ \bibinfo {pages} {5313} (\bibinfo {year}
  {1998})}\BibitemShut {NoStop}%
\bibitem [{\citenamefont {Latora}\ \emph {et~al.}(2001)\citenamefont {Latora},
  \citenamefont {Rapisarda},\ and\ \citenamefont
  {Tsallis}}]{LatoraRapisardaTsallis01}%
  \BibitemOpen
  \bibfield  {author} {\bibinfo {author} {\bibfnamefont {V.}~\bibnamefont
  {Latora}}, \bibinfo {author} {\bibfnamefont {A.}~\bibnamefont {Rapisarda}}, \
  and\ \bibinfo {author} {\bibfnamefont {C.}~\bibnamefont {Tsallis}},\ }\href
  {\doibase 10.1103/PhysRevE.64.056134} {\bibfield  {journal} {\bibinfo
  {journal} {Phys. Rev. E}\ }\textbf {\bibinfo {volume} {64}},\ \bibinfo
  {pages} {056134} (\bibinfo {year} {2001})}\BibitemShut {NoStop}%
\bibitem [{\citenamefont {Yamaguchi}\ \emph {et~al.}(2004)\citenamefont
  {Yamaguchi}, \citenamefont {Barr\'e}, \citenamefont {Bouchet}, \citenamefont
  {Dauxois},\ and\ \citenamefont {Ruffo}}]{Yamaguchi_etal04}%
  \BibitemOpen
  \bibfield  {author} {\bibinfo {author} {\bibfnamefont {Y.}~\bibnamefont
  {Yamaguchi}}, \bibinfo {author} {\bibfnamefont {J.}~\bibnamefont {Barr\'e}},
  \bibinfo {author} {\bibfnamefont {F.}~\bibnamefont {Bouchet}}, \bibinfo
  {author} {\bibfnamefont {T.}~\bibnamefont {Dauxois}}, \ and\ \bibinfo
  {author} {\bibfnamefont {S.}~\bibnamefont {Ruffo}},\ }\href {\doibase
  10.1016/j.physa.2004.01.041} {\bibfield  {journal} {\bibinfo  {journal}
  {Physica A}\ }\textbf {\bibinfo {volume} {337}},\ \bibinfo {pages} {36}
  (\bibinfo {year} {2004})}\BibitemShut {NoStop}%
\bibitem [{\citenamefont {Caglioti}\ and\ \citenamefont
  {Rousset}(2008)}]{CagliotiRousset08}%
  \BibitemOpen
  \bibfield  {author} {\bibinfo {author} {\bibfnamefont {E.}~\bibnamefont
  {Caglioti}}\ and\ \bibinfo {author} {\bibfnamefont {F.}~\bibnamefont
  {Rousset}},\ }\href {\doibase 10.1007/s00205-008-0157-x} {\bibfield
  {journal} {\bibinfo  {journal} {Arch. Ration. Mech. Anal.}\ }\textbf
  {\bibinfo {volume} {190}},\ \bibinfo {pages} {517} (\bibinfo {year}
  {2008})}\BibitemShut {NoStop}%
\bibitem [{\citenamefont {Giansanti}\ \emph {et~al.}(2002)\citenamefont
  {Giansanti}, \citenamefont {Moroni},\ and\ \citenamefont
  {Campa}}]{GiansantiMoroniCampa02}%
  \BibitemOpen
  \bibfield  {author} {\bibinfo {author} {\bibfnamefont {A.}~\bibnamefont
  {Giansanti}}, \bibinfo {author} {\bibfnamefont {D.}~\bibnamefont {Moroni}}, \
  and\ \bibinfo {author} {\bibfnamefont {A.}~\bibnamefont {Campa}},\ }\href
  {\doibase 10.1016/S0960-0779(01)00022-4} {\bibfield  {journal} {\bibinfo
  {journal} {Chaos Soliton Fract.}\ }\textbf {\bibinfo {volume} {13}},\
  \bibinfo {pages} {407} (\bibinfo {year} {2002})}\BibitemShut {NoStop}%
\bibitem [{\citenamefont {McLachlan}\ and\ \citenamefont
  {Atela}(1992)}]{McLachlanAtela92}%
  \BibitemOpen
  \bibfield  {author} {\bibinfo {author} {\bibfnamefont {R.~I.}\ \bibnamefont
  {McLachlan}}\ and\ \bibinfo {author} {\bibfnamefont {P.}~\bibnamefont
  {Atela}},\ }\href {\doibase 10.1088/0951-7715/5/2/011} {\bibfield  {journal}
  {\bibinfo  {journal} {Nonlinearity}\ }\textbf {\bibinfo {volume} {5}},\
  \bibinfo {pages} {541} (\bibinfo {year} {1992})}\BibitemShut {NoStop}%
\bibitem [{\citenamefont {Yamaguchi}(2003)}]{Yamaguchi03}%
  \BibitemOpen
  \bibfield  {author} {\bibinfo {author} {\bibfnamefont {Y.~Y.}\ \bibnamefont
  {Yamaguchi}},\ }\href {\doibase 10.1103/PhysRevE.68.066210} {\bibfield
  {journal} {\bibinfo  {journal} {Phys. Rev. E}\ }\textbf {\bibinfo {volume}
  {68}},\ \bibinfo {pages} {066210} (\bibinfo {year} {2003})}\BibitemShut
  {NoStop}%
\bibitem [{Note1()}]{Note1}%
  \BibitemOpen
  \bibinfo {note} {Care must be taken when performing the limits of large $N$
  and $t$, as the prefactor $\protect \mathcal {N}$ may affect the order of
  these limits.}\BibitemShut {Stop}%
\bibitem [{\citenamefont {Firpo}\ and\ \citenamefont
  {Ruffo}(2001)}]{FirpoRuffo01}%
  \BibitemOpen
  \bibfield  {author} {\bibinfo {author} {\bibfnamefont {M.-C.}\ \bibnamefont
  {Firpo}}\ and\ \bibinfo {author} {\bibfnamefont {S.}~\bibnamefont {Ruffo}},\
  }\href {\doibase 10.1088/0305-4470/34/37/102} {\bibfield  {journal} {\bibinfo
   {journal} {J. Phys. A: Math. Gen.}\ }\textbf {\bibinfo {volume} {34}},\
  \bibinfo {pages} {L511} (\bibinfo {year} {2001})}\BibitemShut {NoStop}%
\bibitem [{\citenamefont {Anteneodo}\ and\ \citenamefont
  {Vallejos}(2001)}]{AnteneodoVallejos01}%
  \BibitemOpen
  \bibfield  {author} {\bibinfo {author} {\bibfnamefont {C.}~\bibnamefont
  {Anteneodo}}\ and\ \bibinfo {author} {\bibfnamefont {R.~O.}\ \bibnamefont
  {Vallejos}},\ }\href {\doibase 10.1103/PhysRevE.65.016210} {\bibfield
  {journal} {\bibinfo  {journal} {Phys. Rev. E}\ }\textbf {\bibinfo {volume}
  {65}},\ \bibinfo {pages} {016210} (\bibinfo {year} {2001})}\BibitemShut
  {NoStop}%
\bibitem [{\citenamefont {Latora}\ and\ \citenamefont
  {Baranger}(1999)}]{LatoraBaranger99}%
  \BibitemOpen
  \bibfield  {author} {\bibinfo {author} {\bibfnamefont {V.}~\bibnamefont
  {Latora}}\ and\ \bibinfo {author} {\bibfnamefont {M.}~\bibnamefont
  {Baranger}},\ }\href {\doibase 10.1103/PhysRevLett.82.520} {\bibfield
  {journal} {\bibinfo  {journal} {Phys. Rev. Lett.}\ }\textbf {\bibinfo
  {volume} {82}},\ \bibinfo {pages} {520} (\bibinfo {year} {1999})}\BibitemShut
  {NoStop}%
\bibitem [{\citenamefont {Hastings}\ and\ \citenamefont
  {Koma}(2006)}]{HastingsKoma06}%
  \BibitemOpen
  \bibfield  {author} {\bibinfo {author} {\bibfnamefont {M.~B.}\ \bibnamefont
  {Hastings}}\ and\ \bibinfo {author} {\bibfnamefont {T.}~\bibnamefont
  {Koma}},\ }\href {\doibase 10.1007/s00220-006-0030-4} {\bibfield  {journal}
  {\bibinfo  {journal} {Commun. Math. Phys.}\ }\textbf {\bibinfo {volume}
  {265}},\ \bibinfo {pages} {781} (\bibinfo {year} {2006})}\BibitemShut
  {NoStop}%
\bibitem [{\citenamefont {Wreszinski}(2010)}]{Wreszinski10}%
  \BibitemOpen
  \bibfield  {author} {\bibinfo {author} {\bibfnamefont {W.~F.}\ \bibnamefont
  {Wreszinski}},\ }\href {\doibase 10.1007/s10955-009-9889-8} {\bibfield
  {journal} {\bibinfo  {journal} {J. Stat. Phys.}\ }\textbf {\bibinfo {volume}
  {138}},\ \bibinfo {pages} {567} (\bibinfo {year} {2010})}\BibitemShut
  {NoStop}%
\bibitem [{Note2()}]{Note2}%
  \BibitemOpen
  \bibinfo {note} {This is similar to what happens in equilibrium statistical
  mechanics, where the threshold at which equilibrium properties switch from
  long-range to short-range behavior is universal in the absence of randomness,
  but can get modified when randomness is present.}\BibitemShut {Stop}%
\bibitem [{\citenamefont {Bachelard}\ \emph {et~al.}(2011)\citenamefont
  {Bachelard}, \citenamefont {Dauxois}, \citenamefont {Ninno}, \citenamefont
  {Ruffo},\ and\ \citenamefont {Staniscia}}]{Bachelard2011}%
  \BibitemOpen
  \bibfield  {author} {\bibinfo {author} {\bibfnamefont {R.}~\bibnamefont
  {Bachelard}}, \bibinfo {author} {\bibfnamefont {T.}~\bibnamefont {Dauxois}},
  \bibinfo {author} {\bibfnamefont {G.~D.}\ \bibnamefont {Ninno}}, \bibinfo
  {author} {\bibfnamefont {S.}~\bibnamefont {Ruffo}}, \ and\ \bibinfo {author}
  {\bibfnamefont {F.}~\bibnamefont {Staniscia}},\ }\href {\doibase
  10.1103/PhysRevE.83.061132} {\bibfield  {journal} {\bibinfo  {journal} {Phys.
  Rev. E}\ }\textbf {\bibinfo {volume} {83}},\ \bibinfo {pages} {061132}
  (\bibinfo {year} {2011})}\BibitemShut {NoStop}%
\bibitem [{\citenamefont {Bouchet}\ and\ \citenamefont
  {Dauxois}(2005)}]{BouchetDauxois05}%
  \BibitemOpen
  \bibfield  {author} {\bibinfo {author} {\bibfnamefont {F.}~\bibnamefont
  {Bouchet}}\ and\ \bibinfo {author} {\bibfnamefont {T.}~\bibnamefont
  {Dauxois}},\ }\href {\doibase 10.1103/PhysRevE.72.045103} {\bibfield
  {journal} {\bibinfo  {journal} {Phys. Rev. E}\ }\textbf {\bibinfo {volume}
  {72}},\ \bibinfo {pages} {045103(R)} (\bibinfo {year} {2005})}\BibitemShut
  {NoStop}%
\end{thebibliography}%

\end{document}